# ON THE POSSIBILITY OF IMPROVING THE ORBITS OF SATELLITES BASED ON OBSERVATIONS OF ISOLATED X-RAY PULSARS

© 2015 **Revnivtsev M.G.*[1], Gadzhily O.E.[1,2], Lutovinov A.A.[1], Molkov S.V.[1], Arefiev V.A.[1], Pavlinsky M.N.[1], Tuchin A.G.[3]**

[1]*Space Research Institute, Moscow, Russia*
[2]*Moscow Institute of Physics and Technology, Dolgoprudnyi, Moscow reg., Russia*
[3]*Keldysh Institute of Applied Mathematics, Moscow, Russia*

At present, there is a great worldwide interest in the development of technologies that allow information about the X-ray emission from pulsating cosmic sources to be used to obtain navigation solutions for deep-space spacecraft. In this paper, we illustrate the technique for determining the spatial position of a spacecraft based on the already existing data from the RXTE X-ray space observatory. We show that the spacecraft position toward the Crab pulsar can be determined using an X-ray detector with an effective area of about 0.6 $m^2$ in the energy range 3-15 keV with an accuracy up to 730 m in a signal integration time of 1000 s. Extending the energy range to 1 keV (the efficiency of the RXTE/PCA spectrometer decreases dramatically at energies below 3 keV) will allow a spacecraft position accuracy of 400-450 m to be achieved at the same effective area and up to 300-350 m when using detectors with an effective area of $\simeq 1$ $m^2$ in the energy range 1-10 keV.

## INTRODUCTION

The most important characteristic of a spacecraft needed for its efficient operation is the precise knowledge of its position in space. The triaxial orientation of the spacecraft can be reconstructed with an ordinary star tracker, but its spatial position and velocity vector are fairly difficult to determine.

For spacecraft in low orbits, the determination of motion parameters is simplified considerably by using the GLONASS and GPS navigation systems (see, e.g., Akim et al. 2009). The motion parameters of spacecraft in flight trajectories to the Moon and planets as well as in the orbits of the lunar and planetary satellites are determined from radio trajectory measurements of the radial velocity and slant distance (see, e.g., Tuchin et al. 2013). The accuracy of determining the motion parameters of a spacecraft depends on many factors (see, e.g., Shishov 2008; Zakhvatkin et al. 2014). However, there is one common problem. Since the radio measurements are the distance and velocity measurements toward the Earth, a good accuracy of determining the spacecraft position and velocity is achieved in this direction and a poor one is achieved in an orthogonal direction. Radio-interferometric measurements based on simultaneous signal registration by two widely spaced antennas are used when there is a need to increase the accuracy in directions orthogonal to those toward the Earth. In this paper, we propose to use measurements of the distance from the reference point to the spacecraft toward a pulsar, which is more efficient than interferometric measurements. Such measurements can be used both to determine the motion parameters at a ground-based center and in an autonomous onboard navigation system.

The idea of using astrophysical objects to obtain a spacecraft navigation solution was put forward almost immediately after the discovery of pulsars, rapidly spinning neutron stars that specify a very stable time scale by their signal (see, e.g., the review of Verbiest et al. 2009, and references therein). Back in 1974, Downs (1974) investigated the possibility of using radio pulsars to solve the problems of autonomous interplanetary navigation. Somewhat later, Chester and Butman (1981) considered the use of Xray pulsars, because instruments with a much smaller size than those in the radio band are required to reliably detect sources in the X-ray band.

*e-mail: revnivtsev@iki.rssi.ru

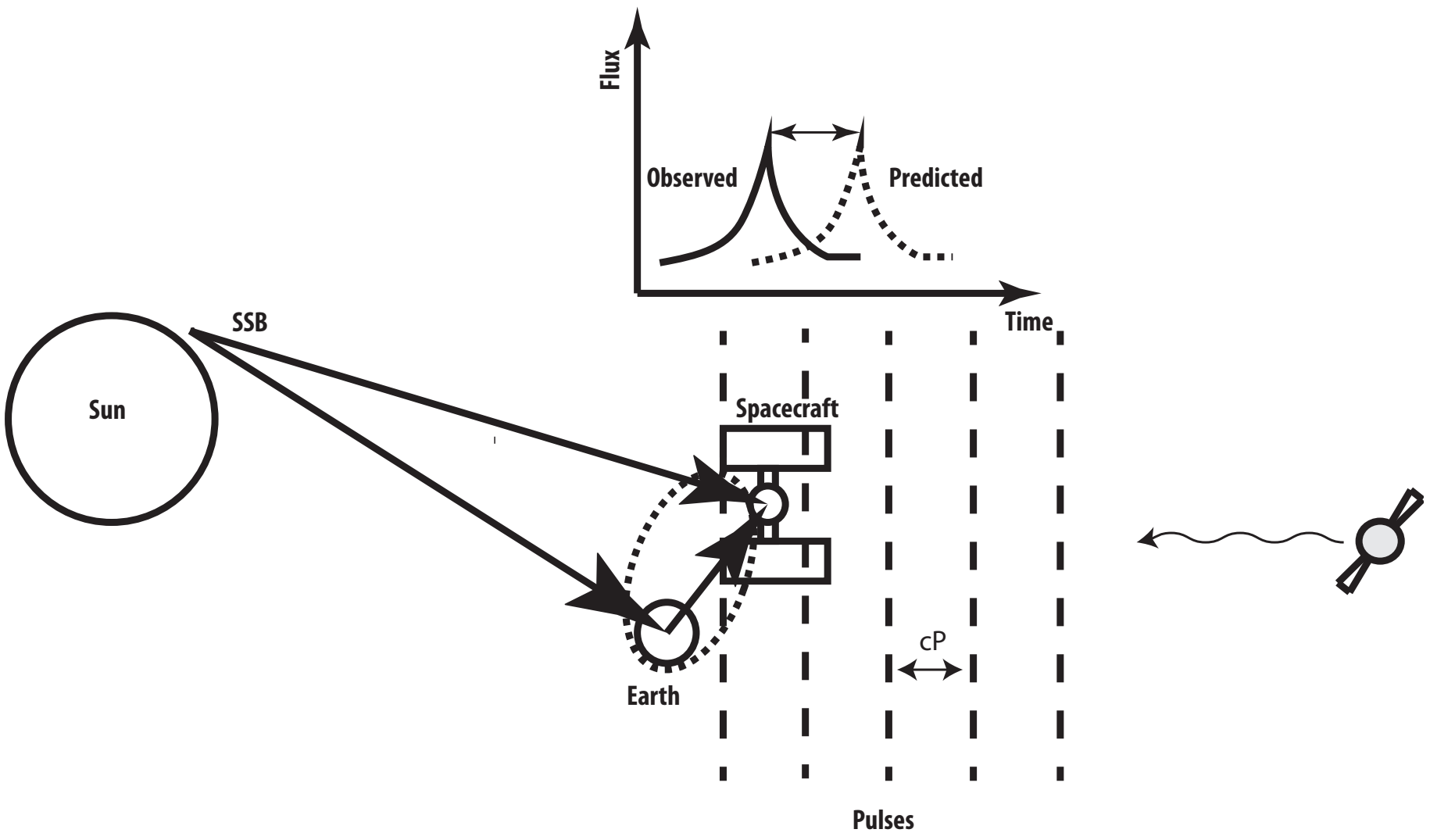

**Fig. 1.** Scheme for determining the spatial position of a spacecraft from observations of an astrophysical pulsating object. The correction to the spacecraft position along the direction toward the astrophysical object is calculated by using information about the time of pulse arrival from such a source at the spacecraft and comparing it with the model.

Since then, the questions of autonomous spacecraft navigation using X-ray pulsars have been considered in a large number of papers and reviews (see, e.g.,cWood 1993; Hanson 1996; Sheikh 2005; Sheikh et al. 2006; Arefiev et al. 2009, 2011; Emadzadeh and Speyer 2011; Becker et al. 2013; Deng et al. 2013; Lutovinov et al. 2014), and a large number of numerical experiments have been carried out. As a result, it has been shown that a spacecraft position accuracy up to several hundred meters can be achieved using an effective collecting area of $\simeq 1$ $m^2$ in the energy range 0.5-10 keV when the signal is integrated for several minutes (Fig. 1).

In this paper, we demonstrate the possibility of determining the spatial position of a spacecraft using data from a real space experiment onboard the RXTE observatory as an example.

## OBSERVATIONS

The RXTE/PCA spectrometer (Bradt et al. 1993) is the X-ray instrument whose measurements are currently best suited to demonstrating the possibility of determining the satellite position based on the signals from X-ray pulsars. The effective area of the PCA spectrometer is $\sim$0.64 $m^2$ at an energy of $\sim$ 6 keV (Jahoda et al. 2006). The time resolution of the instrument is 1 $\mu$s. The absolute timing accuracy is better than 1 $\mu$s (Rots et al. 1998).

In order to check the accuracy of our knowledge of the RXTE satellite orbit, we analyzed the time of arrival (TOA) of pulses from the Crab pulsar. The LHEASOFT V6.15 software package was used to process the data obtained.

To demonstrate the effects arising from the satellite motion in space, we analyzed the "long" observations of the Crab pulsar (pulsar period $P_{\rm Crab} \sim 33.5$ ms) in the period March 22-23, 1997 (ObsID 20804-01-01-...). In these observations, the data were recorded in the format of an event list with a time resolution of 61 $\mu$s (to be more precise, $2^{-14}$ s) in the narrow 2-5 keV energy band. Therefore, the mean event count rate was only 5500 counts/s, which is considerably smaller than that in the total RXTE/PCA range, 13000 counts/s. The average pulse profile for the source in 1000 phase bins was obtained by averaging all data from these observations. An example of how the average profile can be used to determine the TOA shift is shown in Fig.2.

The evolution of the pulse arrival time from the Crab pulsar during $\simeq 5$ h of RXTE observations is shown in Fig.3. To obtain this dependence, we did not recalculate the photon arrival time to the Solar system barycenter. The orbital motion of the satellite around the Earth (with a period of about 5.7 ks) and the long-term change in the pulse arrival time due to the Earth's motion around the Sun are clearly traceable on the presented graph.

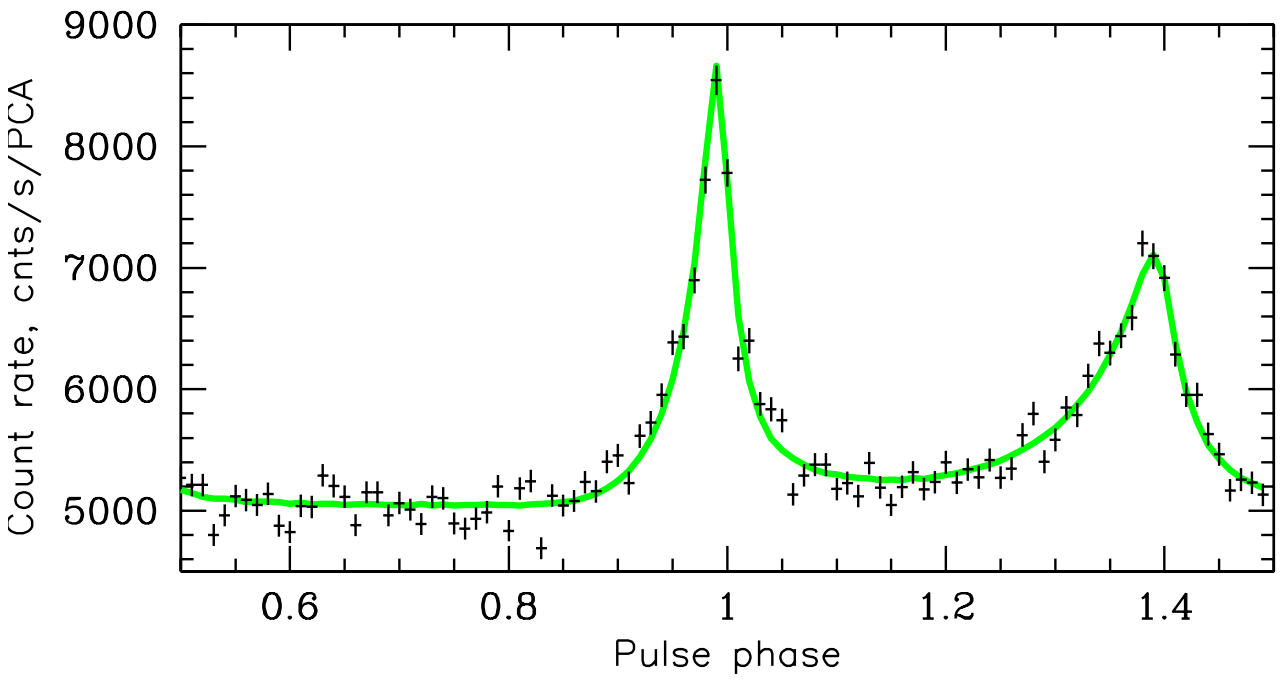

**Fig. 2.** Example of the pulse profile for the Crab pulsar based on RXTE data (PCA, an 2-5 keV energy band, an integration time of 32 s) and its fit by the average profile.

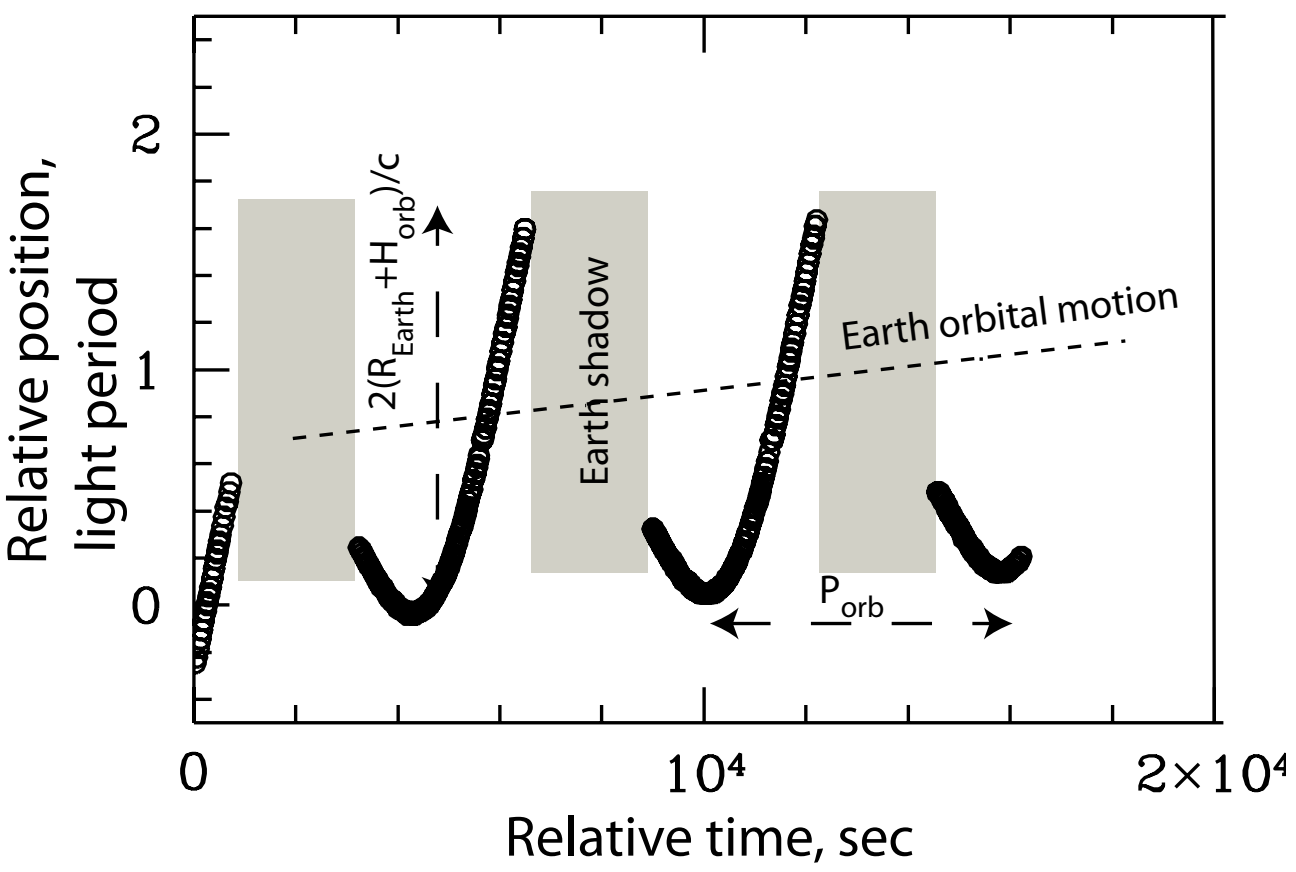

**Fig. 3.** Observed delays of pulse arrival from the Crab pulsar due to the spacecraft motion. The signal integration time is 3 s. The delay time is clearly seen to be modulated by the orbital motion of the satellite around the Earth and the Earth's motion around the Sun. The breaks in observations arise from the fact that the source enters the Earth's shadow.

## RESULTS

To demonstrate the possibility of determining the spacecraft position from observations of the pulsating source in the Crab Nebula, we used the observation performed on February 27, 2003 (Obs ID 70018-01-18-00). All five individual PCA detectors worked in this observation (thus, the instrument's maximum effective area was used), while the event list was generated in the entire operating RXTE/PCA range (effectively 3-15 keV); the time resolution was about 244 $\mu$s (to be more precise, $2^{-12}$ s).

The method of determining the accuracy of our knowledge of the spacecraft position is as follows:

1. The photon arrival time is recalculated to the Solar system barycenter according to the information about the satellite orbit provided in the observation files (FPorbit_Day...)

2. The photons arrived at the spectrometer in 8 s of observations are binned (the fasebin code of the HEASOFT package) in pulse phases according to the pulsar's ephemerides from the library of the Jodrell Bank Observatory (Lyne et al. 1993, http://www.jb.man.ac.uk/~pulsar/crab.html):

$$\Delta\phi(t) = \nu_0(t - t_0) + \frac{1}{2}\dot{\nu}(t - t_0)^2 + \frac{1}{6}\ddot{\nu}(t - t_0)^3$$

where $\nu, \dot{\nu}, \ddot{\nu}$ – are the pulsar spin frequency and its first and second derivatives, respectively (all quantities depend on time), $t_0$ is the reference time of pulse arrival. Here, the pulsar's ephemerides measured in the radio band are assumed to be accurate.

3. The shift of the pulse arrival time relative to the reference time is calculated. The shift is calculated by fitting the observed pulse with its average profile by minimizing the $\chi^2$ value (for the averaging time of 8 s used, the mean signal significance reaches $30\sigma$, and the approximation ofGaussian errors in the count rate is valid).

4. The accuracy of determining the shift of the pulse arrival is a measure of how accurately we know the satellite position in projection onto the direction toward the pulsar. If the pulse arrival time will differ from the reference one after allowance for these effects (the time evolution of the period and the influence of the satellite orbit), then this will mean an error in our knowledge of the spacecraft position. For example, problems in the operation of the GPS equipment were detected in this way during the observations of the Crab pulsar with the *U*SA spectrometer onboard the ARGOS satellite (Wood et al. 1994), which led to errors in the satellite position of $\sim$ 10 km (Sheikh 2005).

Our determination of the pulse arrival time from the Crab pulsar is shown in Fig. 4. In our case, we do not observe any deviations of the TOA from the predicted ones to within $\Delta\phi \sim 7.8 \times 10^{-4}$ $P$, where $P \approx 33.5$ ms is the pulsar period. In a signal integration time of only 8 s, the uncertainty in the pulse shift at the $1\sigma$ level is $\Delta\phi \approx 7.8 \times 10^{-4}$ $P$. Consequently, the spacecraft position accuracy along the direction toward the pulsar can be estimated as $\Delta l = \Delta\phi Pc \approx 7.8$ km.

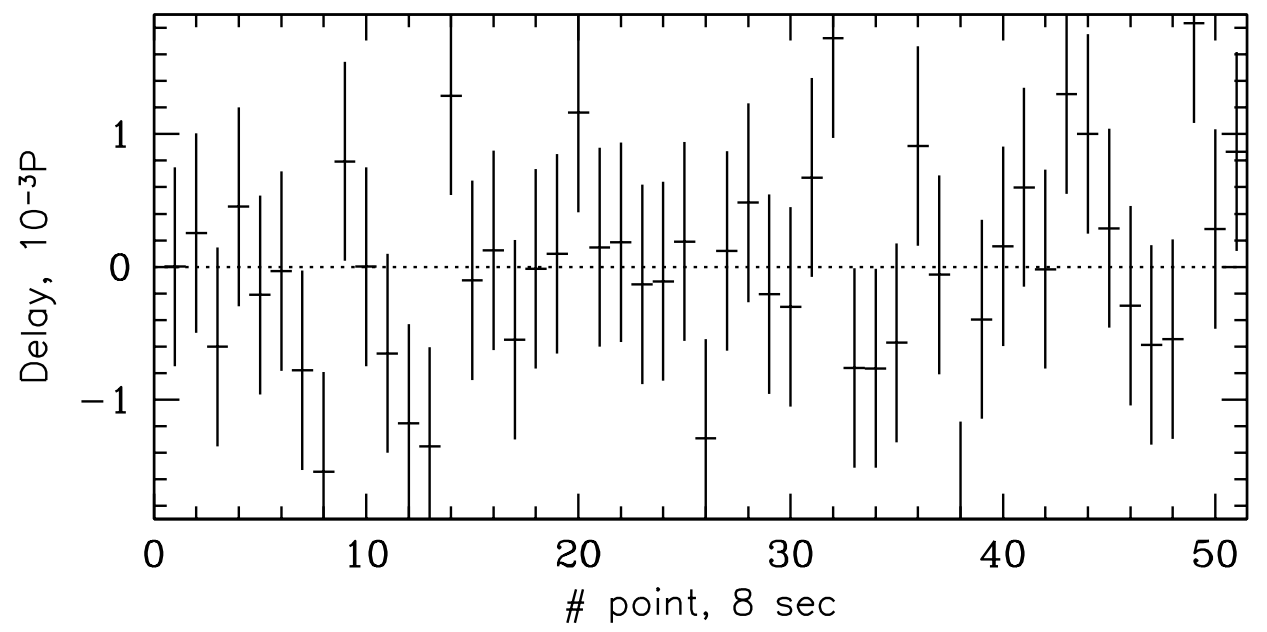

**Fig. 4.** Shift of the time of pulse arrival from the Crab pulsar relative to the model one after allowance for the satellite orbit in the barycentric coordinate system, $10^{-3}P$. A nonzero time of arrival means a shift of the satellite position along the direction toward the pulsar: $\delta l = \delta\phi Pc$, where $P \sim 33.5$ ms is the pulsar period, $c$ is the speed of light.

As the signal integration time increases, the position accuracy improves in accordance with the relation $\Delta l \sim 2.3 \times 10^{-3} cP\ T_c^{-1/2}$, where $T_c$ is the normalized signal integration time in seconds. In particular, the accuracy of the corrections to the spacecraft position for an integration time of $\sim 1000$ s is approximately 730 m.

It should be noted that the RXTE/PCA spectrometer does not register all of the photons arriving from the source. Its effective area decreases rapidly at energies below 3 keV, which excludes most of the photons emitted by the Crab pulsar (see the source's spectrum in Fig. 5); the decrease in the photon flux density at energies below 1 keV is the result of interstellar absorption in the Galaxy; the column density of the matter toward the pulsar is $N_H L \sim 0.4 \times 10^{22}$ cm$^{-2}$ (Weisskopf et al. 2011). There is also a considerable decrease in the detector efficiency at energies above 10 keV. However, this does not affect so strongly the number of registered events compared to energies $< 3$ keV because of the overall drop in the photon flux density at high energies.

For a detector registering $\sim 100\%$ of the photons in the energy range 1-10 keV, the event count rate during observations of the Crab pulsar by an instrument with a collecting area of 0.64 m$^2$ must be $\sim$ 6.3 counts/s/cm$^2$ $\times$ 6400 cm$^2$ $\sim$ 40.5 thousand counts/s (cf. 13 thousand counts/s actually registered by RXTE/PCA).

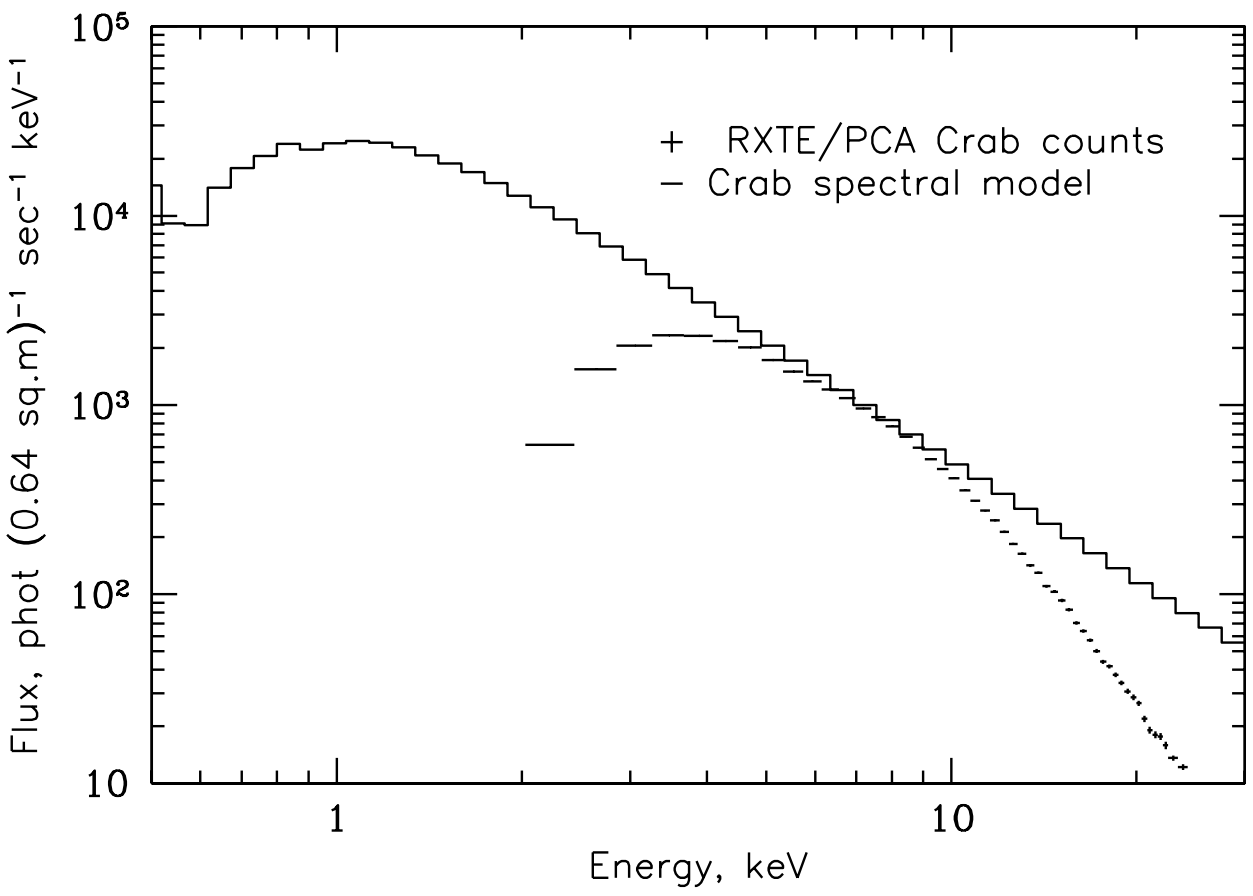

**Fig. 5.** Spectral model of the source in the Crab Nebula recalculated to an effective area of 0.64 m$^2$ (histogram). The spectrum of events measured with the RXTE/PCA spectrometer is represented by the crosses. The ratio of the spectrum of events recorded by the RXTE/PCA spectrometer at energies below 4 keV and above 10 keV to the true spectrum of the source decreases due to a decrease in the efficiency of the PCA gas counter.

Since the accuracy of determining the shift of the pulse arrival time depends on the mean count rate approximately as the square root of the number of counts per second, when using a detector with a high efficiency down to an energy of 1 keV and with the same geometric area as that of RXTE/PCA, the accuracy of determining the corrections to the spacecraft position will be $\Delta l \sim \sqrt{13/40} \times 730$ m $\sim 416$ m. At an effective area of 1 m$^2$ in the energy range 1-10 keV, the event count rate will be $\sim 63$ thousand counts/s, and, consequently, the accuracy of determining the corrections to the spacecraft position will be about 330 m.

## CONCLUSIONS

In this paper, we used the source in the Crab Nebula as an example to demonstrate the possibility of improving the spatial position of a spacecraft by analyzing the signal from X-ray pulsars. We showed that corrections to the satellite position of $\sim 730$ m could be obtained using the RXTE/PCA spectrometer with an effective area of about 0.64 m$^2$ when the signal was integrated for $\sim 1000$ s. The possible accuracy of determining the corrections to the spacecraft position can be improved significantly when using detectors with a high efficiency at energies below 3 keV (the efficiency for the RXTE/PCA spectrometer decreases rapidly toward low energies): up to $\sim 420$ m for a detector with an area of 0.64 $m^2$ and up to $\simeq 330$ m for a detector with an effective area of $\sim 1$ $m^2$ in the energy range 1-10 keV.

*This work was supported by the Russian Science Foundation (project no. 14-12-01287). Authors thanks to V.Astakhov for the translation of the manuscript.*